\documentclass{mem}

\usepackage{natbib}
\usepackage{txfonts}
\usepackage{balance}
\usepackage{graphicx}
\usepackage[a4paper]{hyperref}
\idline{00}{169}

\newcommand{\Md}{M_{\rm d}}

\newcommand{\Mb}{M_{\rm b}}

\newcommand\degrees{^\circ}

\newcommand{\panuc}{\mbox{$\psi_{\rm nuc}$}}

\begin{document}

\title{Long-lived double-barred galaxies in N-body simulations }

\subtitle{}

\author{J. Shen\inst{1,2} \and V. P. Debattista\inst{3}}

\offprints{Juntai Shen; \email{jshen@shao.ac.cn}}
 
\institute{Department of Astronomy, University of Texas, Austin, 
  TX, USA
\and
Current address: Key Laboratory for Research in Galaxies and Cosmology, Shanghai Astronomical Observatory, Chinese Academy of Sciences, 80 Nandan Road, Shanghai 200030, China
\and
  Centre for Astrophysics, University of Central Lancashire, Preston, UK}

\authorrunning{Shen \& Debattista}

\titlerunning{Double-barred galaxies}

\abstract{ 
Many barred galaxies harbor small-scale secondary bars in the
center. The evolution of such double-barred galaxies is still not well
understood, partly because of a lack of realistic N-body models with
which to study them. Here we report the generation of such systems in
the presence of rotating pseudobulges. We demonstrate with high mass
and force resolution collisionless N-body simulations that long-lived
secondary bars can form spontaneously without requiring gas, contrary
to previous claims. We find that secondary bars rotate faster than
primary ones. The rotation is not rigid: the secondary bars pulsate,
with their amplitude and pattern speed oscillating as they rotate
through the primary bars. This self-consistent study supports previous
work based on orbital analysis in the potential of two rigidly
rotating bars. We also characterize the density and kinematics of the
N-body simulations of the double-barred galaxies, compare with
observations to achieve a better understanding of such galaxies.  The
pulsating nature of secondary bars may have important implications for
understanding the central region of double-barred galaxies.}

\maketitle{}
 
\section{Introduction}   

Recent imaging surveys have revealed the frequent existence of nuclear
bars in a large number of barred galaxies, e.g., \citet{erw_spa_02}
found that double-barred galaxies are surprisingly common: at least
one quarter of their sample of 38 early-type optically-barred galaxies
harbor small-scale secondary bars. They found that a typical secondary
bar is about $12\%$ the size of its primary counterpart. The facts
that inner bars are also seen in near-infrared
\citep[e.g.,][]{mul_etal_97,lai_etal_02}, and they are often found in
gas-poor S0s indicate that most of them are stellar structures.
Results from these surveys also show inner bars are at a random angle
relative to the primary bars, implying that they are probably
dynamically independent structures. \citet{shl_etal_89} invoked
multiple nested bars to channel gas inflow into galactic centers to
feed AGN, in a similar fashion as the primary bar drives gas inward.
However, recent work suggests that this mechanism may not be as
efficient as originally hoped \citep[e.g.,][]{mac_etal_02}.

Simulations offer the best way to understand double barred
systems. However, the decoupled nuclear bars that formed in early
simulations did not last long.  For example, the most long-lived
nuclear bar in \citet{fri_mar_93} lasted for less than two turns of
the primary bar, corresponding to about 0.4 Gyr, which is far too
short to explain the observed abundance of nested bars.  Furthermore,
their models usually require substantial amounts of gas to form and
maintain these nuclear bars. \citet{hel_etal_07, hel_etal_07_2}
reported that nested bars form in a quasi-cosmological setting, but
the amplitudes of the bars also seem to weaken rapidly after most of
gas has formed stars \citep{hel_etal_07}. \citet{pet_wil_04} found
that 4 out of 10 double-barred galaxies contain very little molecular
gas in the nuclear region. These clues suggest that large amounts of
molecular gas may not be necessary to maintain central nuclear bars.

On the side of orbital studies, \citet{mac_spa_97,mac_spa_00}
discovered a family of loop orbits that may form building blocks of
long-lived nuclear stellar bars (see also \citealt{mac_ath_07}). Their
studies are very important for understanding double barred galaxies,
but their models are not fully self-consistent, since nested bars in
general cannot rotate rigidly through each other
\citep{lou_ger_88}. So fully self-consistent N-body simulations are
still needed to check if their main results still hold when the
non-rigid nature of the bars is taken into account.

Here we demonstrate that long-lived secondary bars can form in purely
collisionless N-body simulations, when a rotating pseudobulge is
introduced in the model \citep[see also][]{deb_she_07,she_deb_09}. The
nuclear bars in our work are distinctly bars, and do not have a spiral
shape. We show that the behavior of our models are in good agreement
with the loop orbit predictions of \citet{mac_spa_00}. We also analyze
the photometrical and kinematical properties of high resolution
models. Our theoretical results can also be compared to the observed
2D kinematics of some double-barred galaxies, to achieve a better
understanding of the dynamics of the secondary bars.

\section{Model setup}

Our high-resolution simulation consists of a live disk and bulge
component. We do not include a halo component for simplicity, also
because secondary bars are very small-scale phenomena in galactic
centers where visible matter is dominant.  The initial disk has the
exponential surface density profile and Toomre's $Q\sim2.0$. The bulge
was generated using the method of \citet{pre_tom_70} as described in
\citet{deb_sel_00}, where a distribution function is integrated
iteratively in the global potential, until convergence. The bulge has
the mass of $\Mb=0.2\Md$.  The bulge set up this way is un-rotating,
we then give the rotation of the bulge by simply reversing the
negative azimuthal velocities of all bulge particles into the same
positive values. We have checked that systems set up this way are in
very good virial equilibrium.

\section{Results}

\begin{figure}[!ht]
\centerline{ \includegraphics[angle=-90.,width=\hsize]{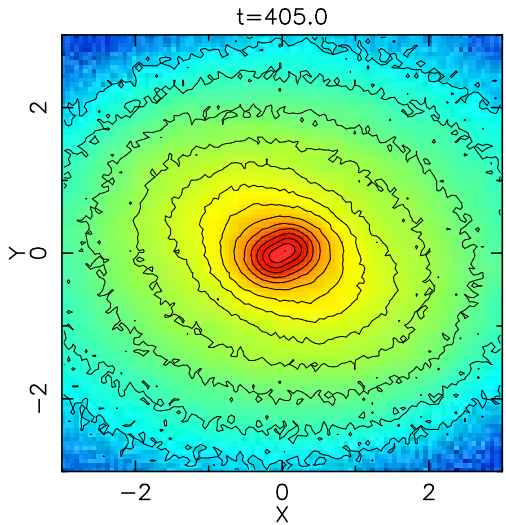} }
\caption{\footnotesize
  The double-barred model at $t=405$ projected to $i = 45\degrees$ and
  $\panuc = 45\degrees$ with all particles shown.  The model bears a
  passing resemblance to NGC 2950.
\label{fig:n2950mod}}
\end{figure}

Fig.~\ref{fig:n2950mod} shows the projected double-barred model (at
$t=405$ when the two bars are nearly perpendicular) with an ordinary
orientation: the system is inclined at $i = 45\degrees$ with the line
of nodes (LON) of $\panuc =45\degrees$ relative to the secondary bar
major axis.  The surface density image and contours resemble many
observed double-barred systems, such as NGC 2950, even though we did
not deliberately set out to match it.

\begin{figure}[!t]
\centerline{
\includegraphics[angle=0.,width=\hsize]{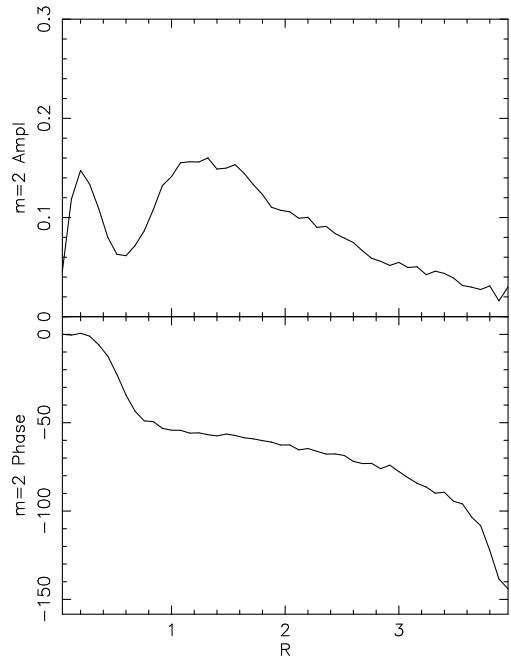}}
\caption{\footnotesize
  The radial variations of the $m=$2 Fourier amplitude and phase of
  all particles for Run D at $t=400$.}
\label{fig:m_2}
\end{figure}

\begin{figure}[!t]
\centerline{
\includegraphics[angle=0.,width=\hsize]{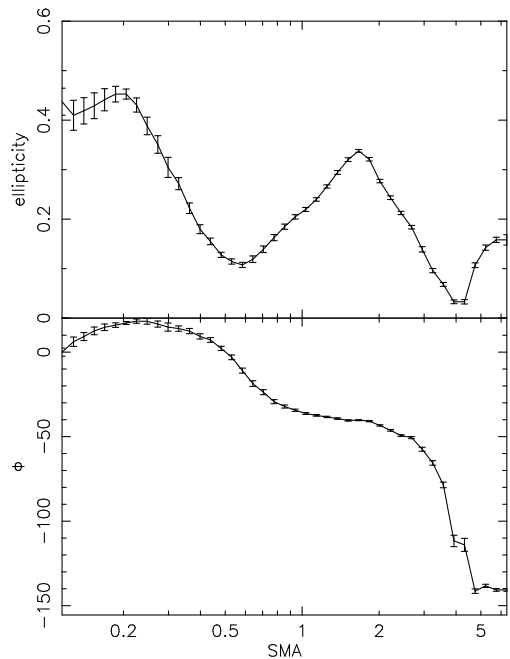}}
\caption{\footnotesize
  Ellipticity and position angle as a function of semi-major axis of
  IRAF-fitted ellipses for Run D at $t=400$.}
\label{fig:iraf}
\end{figure}

Fig.~\ref{fig:m_2} shows radial variations of $m=2$ Fourier amplitude
and phase for run D at $t=400$. Fig.~\ref{fig:iraf} shows the
ellipticity and position angle (PA) profiles of ellipses fitted with
IRAF for the same data as in Fig.~\ref{fig:m_2} (we use log scale for
radius to be consistent with what observers usually adopt). There are
four popular methods for determining the semi-major axis $a_{\rm B}$ of a
bar, as summarized by \citet{one_dub_03} and \citet{erwin_05}. For
convenience, we denote the primary bar as B1 and the secondary bar as
B2:

\begin{itemize}

\item[(1)] the bar end is measured by extrapolating half-way down the
  slope on the $m=2$ amplitude plot (Fig~\ref{fig:m_2}a). We find
  $a_{\rm B1} \sim 2.3$, $a_{\rm B2} \sim 0.4$, the B2/B1 bar length
  ratio is about $\sim 0.17$.

\item[(2)] the bar end is measured when $m=2$ phase deviates from a
  constant by $10^\circ$ (Fig~\ref{fig:m_2}b). We find $a_{\rm B1}
  \sim 2.1$, $a_{\rm B2} \sim 0.4$, the B2/B1 bar length ratio is
  about $\sim 0.19$.

\item[(3)] the bar end is measured at the peak of the fitted
  ellipticity profiles \citep[e.g.,][]{mar_jog_07,men_etal_07}, which
  is shown in Fig~\ref{fig:iraf}a. We find $a_{\rm B1} \sim 1.7$,
  $a_{\rm B2} \sim 0.2$, the B2/B1 bar length ratio is about $\sim
  0.12$.

\item[(4)] the bar end is measured when the PA of fitted ellipses
  deviates from a constant by $10^\circ$ (Fig~\ref{fig:iraf}b). We
  find $a_{\rm B1} \sim 2.3$, $a_{\rm B2} \sim 0.4$, the B2/B1 bar
  length ratio is about $\sim 0.17$.

\end{itemize}

Method 1, 2 and 4 yield consistent values of the bar lengths and
length ratios. We found that method 3 tends to give a lower value of
bar lengths than the other three methods, as shown in
\citet{one_dub_03}. Although these methods have some uncertainties in
measuring the bar lengths, the length ratio of the two bars is in the
range of 0.12 to 0.19 (in particular method 1, 2, and 4 give a
consistent narrow range of 0.17 to 0.19). This result is in good
agreement with the typical observed length ratio of local S2B systems
(median ratio $\sim$ 0.12, see \citealt{erw_spa_02, erwin_04,
  lis_etal_06}).  Note that we expect that the length of the secondary
cannot be too large, otherwise the gravitational torque from the
primary bar will inevitably twist the secondary into alignment if they
rotate at different pattern speeds.

\begin{figure}[!t]
\centerline{
\includegraphics[angle=-90.,width=\hsize]{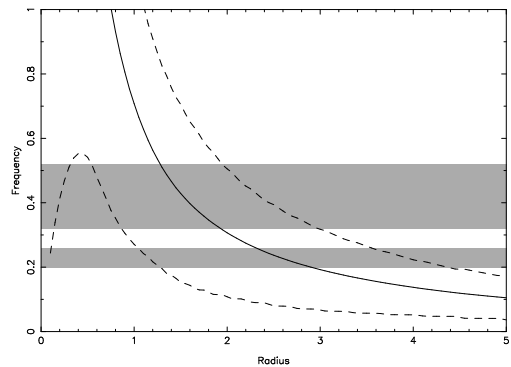}}
\caption{\footnotesize
  Frequencies as a function of radius at around $t=400$ for run D,
  calculated based on the azimuthally averaged gravitational
  attraction. The full-drawn line shows the curve of the circular
  angular frequency $\Omega$ and the dashed curves mark
  $\Omega\pm\kappa/2$, where $\kappa$ is the epicyclic frequency. The
  two shaded bands show the oscillational ranges of the bar pattern
  speeds (the upper band is for the secondary bar and the lower one is
  for the primary). }
\label{fig:spct}
\end{figure}

Fig.~\ref{fig:spct} shows the behavior of the azimuthally averaged
$\Omega$, $\Omega\pm\kappa/2$, and the location of the Lindblad
resonances of the bars at around $t=400$. As shown in
\citet{deb_she_07}, the pattern speeds of the bars, especially that of
the secondary, vary as they rotate through each other: the secondary
bar rotates slower than average when the two bars are perpendicular,
and faster when the bars are parallel. The patten speed bands shown in
Fig.~\ref{fig:spct} reflect such variations. Clearly the pattern speed
of the secondary bar oscillates much more than that of the
primary. The primary bar extends roughly to its CR radius ($\sim2.5$),
consistent with the general expectation and is therefore considered a
fast bar \citep[e.g.,][]{cor_etal_03, deb_wil_04}. The secondary bar rotates
faster than the primary bar. However, the secondary bar is much
shorter than its shortest $R_{\rm CR}$. In addition, even if the
variation of the pattern speed is taken into account, the $R_{\rm CR}$
of the secondary is not very close to the $R_{\rm ILR}$ of the
primary, if we use the same naive definition of $R_{\rm ILR}$ as in
\citet{pfe_nor_90}\footnote{A cautionary note is that the $R_{\rm
    ILR}$ read naively from Fig.~\ref{fig:spct} serves just as a
  visual guide, because the $R_{\rm ILR}$ determined this way is
  reliable only for weak bars, and is questionable for our strong bars
  \citep[e.g.,][]{vanalb_san_82}.}. This is inconsistent with the
CR-ILR coupling proposed to be a requirement for making secondary bars
\citep[e.g.,][]{pfe_nor_90,fri_mar_93}.

We also analyzed the line-of-sight velocity distribution (LOSVD) by
measuring the mean velocity $\overline{v}$ and velocity dispersion
$\sigma$. Departures from a Gaussian distribution are parametrized by
Gauss-Hermite moments \citep{gerhar_93, van_fra_93,ben_etal_94}. The
second order term in such an expansion is related to the dispersion.
$h_3$ measures deviations that are asymmetric about the mean, while
$h_4$ measures the lowest order symmetric deviations from Gaussian
(negative for a `flat-top' distribution, and positive for a more
peaked one).

\begin{figure*}[!t]
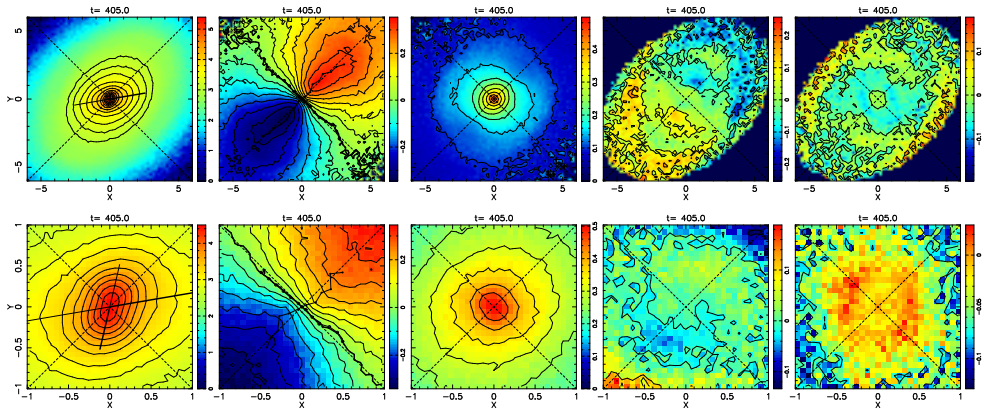

\centerline{
\includegraphics[angle=-90.,width=0.95\hsize]{shen_f05a.ps}}
\vspace{0.01\hsize}
\centerline{
\includegraphics[angle=-90.,width=0.95\hsize]{shen_f05b.ps}}
\caption{\footnotesize
  Photometrical and kinematic maps of our double-barred model. For
  each row from left to right are the projected surface density, mean
  velocities, velocity dispersion, $h_3$, and $h_4$ maps. (a): Row 1,
  the double-barred model at $t=405$ inclined at $45^\circ$ with the
  LON of $45^\circ$ relative to the primary bar major axis; (b): Row
  2, close-up view of (a). One of the dashed lines represents the line
  of nodes ($45^\circ$), while the other dashed is the anti-LOS
  ($135^\circ$).  }
\label{fig:kinemap}
\end{figure*}

The most striking feature in Fig.~\ref{fig:kinemap} is that the twist
of the kinematic minor axis (i.e., $v_{\rm los}=0$) in the secondary
bar region is weak (see the mean velocity maps in
Fig.~\ref{fig:kinemap}a, \ref{fig:kinemap}b).  The kinematic minor
axis is almost perpendicular to the inclination axis, although there
is a small but noticeable twisted pinch near the kinematic minor axis
in the nuclear region. The weak central twist is mainly due to the
relatively large velocity dispersion, especially in the central region
(likewise at $t=20$ when only the small nuclear bar exists, the
stellar twist is stronger than at $t=405$, but still quite small
compared to the expected twist in gaseous kinematics).  On the other
hand, the twist of the kinematic {\it major} axis is more prominent in
the central region.  \citet{moi_etal_04} found the stellar kinematic
minor axis hardly twists from the PA of the disk in their sample with
the most reliable kinematics, leading them to question whether nuclear
photometric isophotal twists represent {\it bona fide} dynamically
decoupled secondary bars. We demonstrate that an authentic decoupled
secondary bar may indeed produce a very weak twist of the kinematic
minor axis in the stellar velocity field. So a central stellar
velocity map without a strong twist as in \citet{moi_etal_04} does not
necessarily exclude the existence of a decoupled nuclear bar.

\section{Conclusions}

We have analyzed the photometrical and kinematical properties of our
high resolution models, and contrasted them when with or without a
secondary bar. This study also compared the simulated secondary bars
with observations.

In general the shape of secondary bars in our models is reasonable
compared to observed ones. The length ratio of two bars, determined by
various methods, is in the range of 0.12 to 0.19, in good agreement
with \citet{erw_spa_02, erw_spa_03}. The primary extends roughly to
its corotation radius, and therefore fits the definition of a fast bar
(see for example \citealt{agu_etal_03}). Although the secondary bar
rotates more rapidly than the primary, its semi-major axis is much
shorter than its corotation radius, even if we take the oscillation of
the bar patterns speeds into account. We did not find evidence of
CR-ILR coupling \citep[e.g.,][]{pfe_nor_90,fri_mar_93} in our models.

We find that the central twist of kinematic axes is quite weak even if
a secondary bar is present, due to the relatively large velocity
dispersion of stars in the central region. This is consistent with the
2D stellar kinematics of secondary bars studied in
\citet{moi_etal_04}. We do not find a $\sigma$ (velocity dispersion)
drop for our secondary bar model. It is more likely that
$\sigma$-drops are just the signature of newly-formed stars, and it is
not necessarily a unique feature of double-barred systems.

The general agreement between our simulations and observations of
double barred galaxies gives us confidence that the simulations are
capturing the same dynamics as in nature.  This is especially
remarkable because secondary bars are not merely scaled down versions
of primary bars, but have distinctly different kinematic properties.
In the absence of self-consistent simulations, earlier orbit-based
models could not directly confront the challenge from observations
which found such differences.  This demonstrates the advantage of
finally being able to simulate stellar double-barred galaxies, which
had been puzzling for so long.

\begin{acknowledgements}
JS acknowledges support from a Harlan J. Smith fellowship of
McDonald Observatory of UT Austin.  VPD was supported by a Brooks
Prize Fellowship at the University of Washington and received partial
support from NSF ITR grant PHY-0205413.
\end{acknowledgements}


\bibliographystyle{aa}

\end{document}